%% file: templateArxiv.tex
\documentclass{article}

\usepackage{PRIMEarxiv}

\usepackage[utf8]{inputenc} 
\usepackage[T1]{fontenc}    
\usepackage{hyperref}       
\usepackage{url}            
\usepackage{booktabs}       
\usepackage{amsfonts}       
\usepackage{nicefrac}       
\usepackage{microtype}      
\usepackage{lipsum}
\usepackage[ruled,linesnumbered,noend]{algorithm2e}
\usepackage[noend]{algorithmic}
\usepackage{fancyhdr}       
\usepackage{graphicx}       
\usepackage{xcolor}
\usepackage{listings}
\usepackage[numbers]{natbib}
\graphicspath{{media/}}     

\title{CONTRACTFIX: A Framework for Automatically
Fixing Vulnerabilities in Smart Contracts
\thanks{\textit{\underline{Citation}}: 
\textbf{Authors. Title. Pages.... DOI:000000/11111.}} 
}

\author{
  Pengcheng Fang \\
  Case Western Reserve University\\
  \texttt{pxf109@case.edu} \\
   \And
  Peng Gao \\
  Virginia Tech \\
  \texttt{penggao@vt.edu} \\
  \AND
  Yun Peng \\
  The Chinese University of Hong Kong \\
  \texttt{ypeng@cse.cuhk.edu.hk} \\
  \And
  Qingzhao Zhang \\
  Shanghai Jiao Tong University \\
  \texttt{fszqz001@sjtu.edu.cn} \\
  \And
  Tao Xie \\
  Peking University \\
  \texttt{taoxie@pku.edu.cn} \\
  \And
  Dawn Song \\
  UC Berkeley \\
  \texttt{dawnsong@berkeley.edu} \\
  \And
  Prateek Mittal \\
  Princeton Univ. \\
  \texttt{pmittal@princeton.edu} \\
  \And
  Sanjeev R. Kulkarni \\
  Princeton Univ. \\
  \texttt{kulkarni@princeton.edu} \\
  \And
  Zhuotao Liu \\
  Tsinghua University \\
  \texttt{zhuotaoliu@tsinghua.edu.cn} \\
  \And
  Xusheng Xiao \\
  Arizona State University \\
  \texttt{xusheng.xiao@asu.edu}
}
\begin{document}
\maketitle

\begin{abstract}
The increased adoption of smart contracts in many industries has made them an attractive target for cybercriminals, leading to millions of dollars in losses.
Thus, deploying smart contracts with detected vulnerabilities (known to developers) are not acceptable, and fixing all the detected vulnerabilities is needed, which incurs high manual labor cost without effective tool support.

To fill this need, in this paper, we propose ContractFix, a novel framework that automatically generates security patches for vulnerable smart contracts. 
 ContractFix is a general framework that can incorporate different fix patterns for different types of vulnerabilities. Users can use it as a security ``fix-it'' tool that automatically applies patches and verifies the patched contracts before deploying the contracts.
To address the unique challenges in fixing smart contract vulnerabilities, given an input smart contract, ContractFix conducts our proposed ensemble identification based on multiple static verification tools  to identify  vulnerabilities that are amenable for automatic fix. 
Then, ContractFix generates patches using template-based fix patterns, and conducts program analysis (program dependency computation and pointer analysis) for smart contracts to accurately infer and populate the parameter values for the fix patterns.
Finally, ContractFix performs static verification that guarantees the patched contract to be  free of vulnerabilities.  
Our evaluations on $144$ real vulnerable contracts demonstrate that ContractFix can successfully fix $94\%$ of the detected vulnerabilities ($565$ out of $601$) and preserve the expected behaviors of the smart contracts. 
\end{abstract}


\section{Introduction}
As a paradigmatic application of blockchain~\cite{nakamotobitcoin}, smart contracts enable the creation of decentralized general-purpose applications and have received wide adoption~\cite{smartcontract,blockchain-finance,blockchain-casino,blockchain-identity}.
While the correct execution of smart contracts is enforced by the consensus protocol of blockchain, it is challenging to create smart contracts without security vulnerabilities, partly due to the lack of security knowledge by developers in the new ecosystem of smart contract languages (e.g. Solidity~\cite{solidity}) and platforms (e.g. permissionless blockchains such as Ethereum~\cite{smartcontract,Ethereum}). 
Over the past few years, the blockchain community witnessed a number of critical vulnerabilities in smart contracts being exploited by attackers, leading to millions of dollars in losses~\cite{King,frozen,read,atzei2017survey,dao,parity}.
For example, the reentrancy attack on TheDAO contract~\cite{thedao} in 2016 resulted in \$50M worth of Ether being stolen~\cite{dao,reentrybug}.

\begin{figure}[t]
    \centering
    \includegraphics[width=0.45\textwidth]{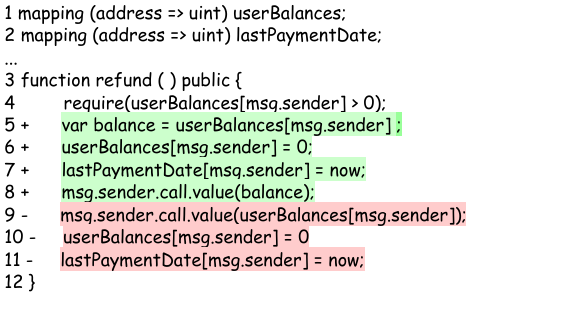}
    \caption{Example patch for \textit{Reentrancy} vulnerability}
    \label{fig:reentrancyExampleDUP}
\end{figure}

Despite considerable research efforts~\cite{luu2016making,mythril,securify,maian,zeus,rodler2018sereum,nguyen2021sguard} of tool support for detecting vulnerabilities in smart contracts, fixing these vulnerabilities is highly critical (yet lacking of effective tool support) for two main reasons. 
First,  unlike other software applications that can be released with known bugs~\cite{avoidbugs}, fixing all detected vulnerabilities before deployment is needed, thus incurring high manual labor cost without effective tool support. 
Such need is partly due to that smart contracts are immutable after deployment; in other words, deploying smart contracts with detected vulnerabilities (known to developers) is not acceptable.
Second, manually fixing a smart contract with multiple detected vulnerabilities is often challenging, and many smart contracts are found to have multiple vulnerabilities (see Sec \ref{subsec:study}). 
For example, the best practice to avoid reentrancy vulnerability is to ensure all internal state changes are performed before the external call is executed (i.e. the Checks-Effects-Interactions pattern)~\cite{smartcontract-practice,reentrybug}. 
Hence, the patch for \emph{reentrancy vulnerability} requires (1) reordering multiple statements to ensure that all updates to contract state variables occur before the external call, and 
(2) creating temporary variables to store the values of these state variables for eliminating data dependencies on the external call (see Fig.~\ref{fig:reentrancyExampleDUP}).

Although various existing techniques of automated program repair~\cite{le2011genprog,nguyen2013semfix,mechtaev2016angelix,long2016automatic,testrepair} can automatically generate patches to fix the given program's bugs, 
these techniques are often not applicable to effectively fix vulnerabilities for smart contracts for two main reasons.  
First, applying existing repair techniques to repair contract vulnerabilities typically requires a  comprehensive test suite to assure that all detected vulnerabilities are fixed and no side effects are introduced by the generated patch.
Previous work~\cite{staticrepair} shows that it is highly difficult to create a comprehensive test suite that can defend against all types of exploits.  
Second, applying existing search-based repair techniques~\cite{genprog,nguyen2013semfix,mechtaev2016angelix,long2016automatic,testrepair,AutoBinary} (being mainstream ones) to fix contract vulnerabilities fails to generate patches for some important types of  contract vulnerabilities. 
These techniques explore the search space of repairs based on syntactic mutators, by leveraging search algorithms such as genetic programming or random search.
However, the strategies of these techniques are mostly adding conditional checks or replacing a statement with another existing statement, which is insufficient for fixing contract vulnerabilities that require temporary variable creations and statement reordering (e.g. fixing reentrancy vulnerabilities).
Although one can simply adapt these techniques to include more complex fixing strategies, doing so  tend to (1) result in an exponential expansion of the search space~\cite{genprog,aprchallenges,aprcontext}, inducing patch-generation ineffectiveness.

To effectively fix vulnerabilities for smart contracts, in this paper, we propose ContractFix, that (1) automatically detects vulnerabilities in a smart contract, (2) applies patches to the multiple detected vulnerabilities, and (3) verifies the patched contract before the contract deployment.
\textit{\textbf{This is the first end-to-end framework that ensembles detection, patching, and verification to fully automate the process of fixing vulnerabilities}}.
In particular, ContractFix is built upon our novel program analysis infrastructure that is specially optimized for Solidity, enabling ContractFix to \textit{support fix strategies for different vulnerabilities and verify the correctness of the patched contracts}. 
In our work, the current instantiation of ContractFix focuses on fixing four major types of vulnerabilities: 
(1) \textit{Reentrancy}, which allows the execution to re-enter a non-recursive function before its termination;
(2) \textit{MissingInputValidation}, which uses default values for function arguments;
(3) \textit{LockedEther}, which  relies on other libraries to transfer ethers; 
(4) \textit{UnhandledException}, which mishandles the raised exceptions. 

ContractFix is powered by three innovative designs. 
First, to avoid wasting later high cost of searching for patches of detected vulnerabilities being false positives or not amenable for automatic fix (e.g. handling external method calls without source code), 
ContractFix \emph{synergistically combines multiple static verification tools~\cite{securify,feist2019slither,smartcheck} with post-processing}. In particular, ContractFix first applies these static verification tools to detect vulnerability candidates and adopts majority voting~\cite{majorityvoting} to determine which candidates are more likely to be real vulnerabilities rather than false-positive ones.  
ContractFix then conducts post-processing to extract the required information from the reported vulnerabilities (e.g. identifying the types of the data dependencies for reentrancy vulnerabilities in Sec.~\ref{subsec:methodlevel}) and filter out candidates that are not amenable for automatic fix.

Second, to address the space explosion during the search for target patches and preserve expected contract behaviors, ContractFix generates patches using template-based fix patterns~\cite{templaterepair},  conducting  \textit{static program analysis} to accurately infer variable values from the contract program under analysis without the need for searching a huge repair space.  
Most smart contracts restrain the uses of references in the language level (e.g. Solidity limits references to specific types),  enabling our static analysis techniques to compute precise program dependencies for generating  complex patches such as moving statements without violating data dependency constraints (Sec.\ref{subsubsec:programanalysis}).
Particularly, our program analysis allows ContractFix to support fix patterns with different performance overheads. 
For example, ContractFix supports both adding global locks and moving statements to fix reentrancy vulnerabilities, and prefers moving statements as the resulting program requires much less gas cost ($5$ v.s. $25000$).

Third, ContractFix reapplies the static verification techniques used to detect the vulnerabilities on the patched smart contract, and verifies that the detected security vulnerabilities are eliminated in the patched smart contract.
In this way, not only our template-based fix patterns with program analysis support can guarantee that the patched smart contract  preserves the expected contract behaviors, the static verification techniques also ensure the elimination of the patched vulnerabilities.

This paper makes the following major contributions:
\begin{itemize}
    \item We propose a novel framework, named ContractFix, that is the \textbf{first work} to (1) leverage the synergy of multiple static verification tools to detect vulnerabilities in a smart contract, (2) generate source code patches for the contract, and (3) performs static verification to verify the patched contract, ensuring the elimination of the vulnerabilities and preserving the expected contract behaviors. 
    \item We propose a novel set of program analysis techniques that extract variable values from smart contracts to generate patches based on the fix patterns for four major types of vulnerabilities. 
    \item We conduct an evaluation on $50$ contracts ($20,510$ LOC) selected from a widely used dataset~\cite{ghaleb2020effective} of smart contracts with injected vulnerabilities and $94$ contracts ($120,894$ LOC) selected among $4,940$ real smart contracts with the largest number of transactions from Etherscan~\cite{Etherscan}. The results show that the majority voting scheme is highly precise in detecting vulnerabilities, and ContractFix changes $7.97$ lines on average to successfully generate patches for $565$  out of $601$ vulnerabilities, \emph{achieving a high success rate ($>94\%$)}.
    \item We crawl $\sim125,000$ transactions from Etherscan~\cite{Etherscan} and replay these transactions on both the original contracts and the patched contracts. The results show that the patched contracts preserve the original contract functionalities, and the increases of the gas caused by the extra security checks are negligible ($\sim\$0.000027$). 
    \item We make a prototype implementation of ContractFix and the evaluation results publicly available~\cite{projectweb}.
\end{itemize}

\section{Background and Empirical Study}
\label{sec:background}

\subsection{Smart Contract and Ethereum}
The very first blockchain, Bitcoin~\cite{nakamotobitcoin}, which supports limited scripting~\cite{bitcoinscript} for its transactions, can already run simple smart contracts such as freezing funds until a time stamp in the future~\cite{bitcoinscript} and decentralized lotteries~\cite{andrychowicz2014secure}.

Ethereum~\cite{Ethereum} and other blockchains (e.g.  Hyperledger~\cite{hyperledger} and Corda~\cite{corda}) support general-purpose computation for smart contracts, and thus it is far less complicated to build a much wider range of decentralized applications (Dapps).
In Ethereum, the Ethereum Virtual Machine (EVM) is a virtual machine designed as the runtime environment for smart contracts.

\subsection{Vulnerabilities in Smart Contracts}
Recently, an increasing number of high-profile attacks resulting in huge financial losses have been reported.
We next illustrate a list of representative vulnerabilities~\cite{atzei2016survey}.

\textbf{Reentrancy} 
In July 2016~\cite{dao}, a fault in TheDAO contract allowed an attacker to steal \$50M. The atomicity and sequentiality of transactions may make developers believe that it is impossible to re-enter a non-recursive function before its termination. 
However, this belief is not always true for smart contracts. 

\begin{figure}[t]
    \centering
    \includegraphics[width=0.45\textwidth]{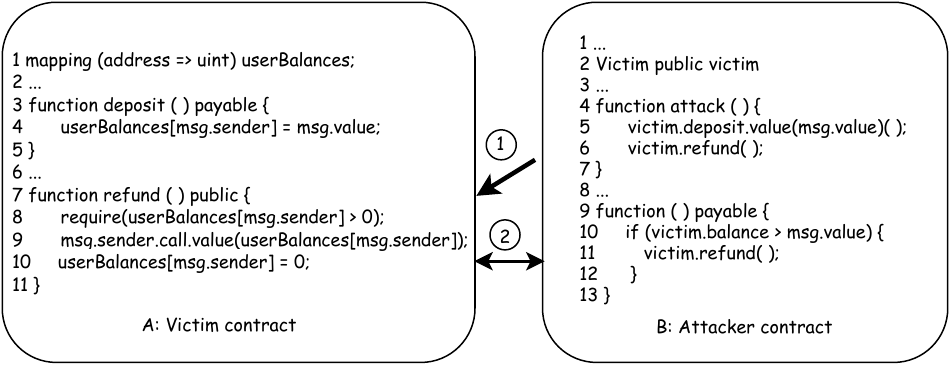}
    \caption{An exploit of \textit{Reentrancy} vulnerability}
    \label{fig:reentrancyAttack}
\end{figure}

Fig.~\ref{fig:reentrancyAttack} shows an exploit of the \textit{Reentrancy} vulnerability.
First, the \textit{attack()} function in the attacker contract is called, causing to  deposit some ethers in the victim contract and then invoke the victim's vulnerable \textit{refund()} function.
Then, the \textit{refund()} function sends the deposited ethers to the attacker contract (Line 9 in A), also triggering the unnamed fallback function in the attack contract (Line 9 in B). 
Next, the fallback function again calls the \textit{refund()} function in the victim contract (Line 11 in B).
Since the victim contract updates the \textit{userBalances} variable (Line 10 in A) after the ether transfer call, \textit{userBalances} remains unchanged when the attacker re-enters the \textit{refund()} function, and thus the balance check (Line 8 in A) can still be passed.
As a consequence, the attacker can repeatedly siphon off ethers from the victim contract and exhaust its balance.

\begin{figure}[t]
    \centering
    \includegraphics[width=0.45\textwidth]{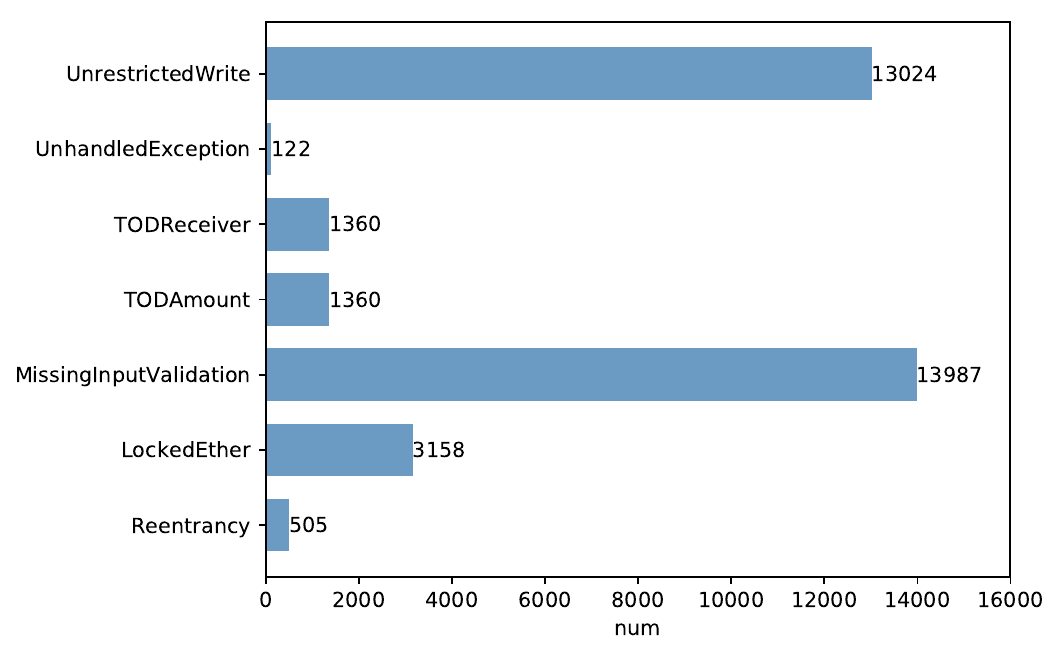}
        \vspace{-6pt}   
    \caption{Vulnerabilities detected by Securify for 4,640 smart contracts collected from Etherscan}
    \label{fig:distribution}
        \vspace{-5pt}   
\end{figure}

\begin{figure*}[t]
    \centering
    \includegraphics[width=0.95\textwidth]{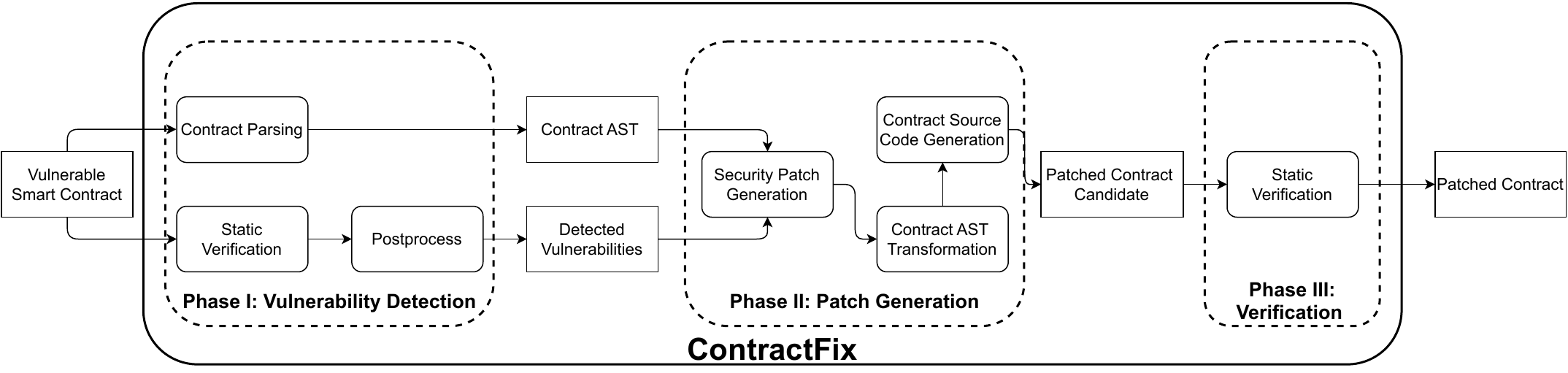}
    \caption{The Architecture of ContractFix}
    \label{fig:architecture}
\end{figure*}

\textbf{Locked Ether}
In 2017, a vulnerable contract led to the frozen of million dollars~\cite{frozen}. 
The reason is that this contract relies on another library contract to withdraw its funds (using \textit{delegatecall}). 
Unfortunately, a user accidentally removed the library contract from the blockchain (using the \textit{kill} instruction), and thus the funds in the wallet contract could not be extracted anymore.

\textbf{Unhandled Exception} 
In Solidity, there are multiple situations where an exception may be raised. 
Unhandled exceptions can affect the security of smart contracts. 
In February 2016, a vulnerable contract~\cite{King} forced the owner to ask the users not to send ether to the owner because of an unhandled exception in the \textit{call} instruction. 

\subsection{Empirical Study}
\label{subsec:study}

As static-verification-based solutions for vulnerability detection use different security properties to detect different types of vulnerabilities, we need to study the  prevalence of vulnerabilities so that ContractFix's  patch generation strategies can target the most effective properties.
Fig.~\ref{fig:distribution} shows the vulnerability distribution obtained by applying Securify on a set of $4,640$ smart contracts (with the most transactions) collected from Etherscan.
In summary, there are $33,516$ vulnerabilities and each contract contains $7.22$ vulnerabilities on average. 
The results show that vulnerabilities are commonly found in smart contracts and multiple vulnerabilities may often exist in one contract, making  manual fixing labor intensive and error prone.
This observation motivates the design of automated solutions to generate patches to fix vulnerabilities.

Based on the vulnerability distribution in Fig.~\ref{fig:distribution}, we select the types of vulnerabilities to include in ContractFix's fixing scope.
The most common types of vulnerabilities are \textit{MissingInputValidation} and \textit{UnrestrictedWrite} (count $>13,000$). 
As \textit{MissingInputValidation} can be fixed via source code transformation, ContractFix includes it in its fixing scope.
For \textit{UnrestrictedWrite}, the security property used to detect the vulnerabilities is too strict, and most of the detected vulnerabilities are false positives.
Thus, ContractFix excludes \textit{UnrestrictedWrite}.

Such uncertainty is inherent in blockchain execution platforms and cannot be fixed by modifying the smart contract source code.
Fixing them needs to change the operational semantics of Ethereum, requiring  all the clients in the Ethereum network to upgrade.

As doing so is not a practical solution, we exclude these types of vulnerabilities from ContractFix's fixing scope.
Furthermore, we include \textit{Reentrancy}, \textit{LockedEther}, and \textit{UnhandledException}.
These types of vulnerabilities are commonly found in smart contracts and their fixing strategies are different from each other, 
making them good candidates to demonstrate the effectiveness of ContractFix in both simple and complex patches. 
The four types of vulnerabilities that ContractFix focuses on account for $53.0\%$ of the total vulnerabilities.

\section{Design Of ContractFix}
\label{sec:approach}
\subsection{Phase I: Vulnerability Detection}
\label{subsec:detection}
In this phase, ContractFix first conducts static verification to detect vulnerability candidates. 
Based on our empirical study (Sec.~\ref{subsec:study}), ContractFix conducts static verification that checks the security properties for identifying four types of vulnerabilities: \textit{Reentrancy}, \textit{MissingInputValidation}, \textit{LockedEther}, \textit{UnhandledException}.
Static verification tools~\cite{securify,luu2016making,mythril} adopt over-approximation analysis, which may produce false-positive violations. 
To address this issue, ContractFix combines three static verification tools: \textit{Securify}~\cite{securify}, \textit{Slither}~\cite{feist2019slither}, and \textit{Smartcheck}~\cite{smartcheck} to detect vulnerabilities, 
and leverages the majority voting mechanism to improve the detection accuracy.

As some of the detected vulnerabilities are not amenable for automatic fix (e.g. handling external method calls without source code), ContractFix focuses on the detected vulnerabilities that have severe security impacts based on our motivating study (Sec~\ref{sec:background}) and are amenable for automatic fix:

\begin{itemize}
    \item \textit{Reentrancy}: 
    These vulnerabilities can be detected by \textit{Slither} and \textit{Securify}.
    However, for some detected vulnerabilities, the return value of external function call is used to control whether to update the state variables.
    As it is almost impossible to verify the behavior of external function calls, ContractFix cannot generate a patch properly.
    Also, some updates of the state variables depend on timestamps, and any patch that moves the updates will cause semantics changes. 
    Thus, ContractFix ignores the reentrancy vulnerabilities caused by external function calls.
    
    \item \textit{MissingInputValidation}:
    This type of vulnerability can be detected by \textit{Securify}.
    Except for function arguments of the \textit{address} type, function arguments of other data types (e.g. integers) can have a wide range of values and it requires dynamic analysis to determine the runtime values.
    Thus, ContractFix only fixes the \textit{MissingInputValidation} vulnerabilities that concern about the \textit{address} type arguments.
    
    \item \textit{LockedEther}: 
    This type of vulnerability can be detected by both \textit{Slither}, \textit{Securify}, and \textit{Smartcheck}.
    For some contracts, developers use them as the library, which is not assumed to receive ether. 
    Thus, for these contracts, it is not reasonable for them to have a function that can send out Ether. 
    Thus, ContractFix will not fix the \textit{LockedEther} violations for these contracts.
    
    \item \textit{UnhandledException}: 
    This type of vulnerability can be detected by \textit{Slither}, \textit{Securify}, and \textit{Smartcheck}.
    When developers assign the return value of the Ether transfer function \textit{send()} to a variable, they often provide more code to handle the exceptions.
    Thus, ContractFix fixes the violations where developer does not process the return value of \textit{send()}.
\end{itemize}

To identify these types of vulnerabilities that are amenable for automatic fix,
ContractFix performs post-processing on the reported vulnerabilities based on the syntactic analysis on the AST and intra-procedural control and data flow analysis.
For example, for a reported reentrancy vulnerability, detecting whether an external function call is used to control the execution of a state variable update will require both control and data flow analysis.

\subsection{Phase II: Patch Generation}
\label{subsec:phase2}

ContractFix performs static program analysis to extract the context information of the detected vulnerabilities, and supports fix patterns in three granularity levels: \textit{statement level}, \textit{method level}, and \textit{contract level}.
Tab.~\ref{tab:patterns} summarizes the fix patterns supported by ContractFix. 
We can see that ContractFix can support a wide range of fix patterns, while existing work often supports one or two patterns~\cite{reentrybug,nguyen2021sguard}.
For example, sGUARD~\cite{nguyen2021sguard} supports only adding locks for fixing reentrancy vulnerabilities, while ContractFix additionally supports the fix pattern by reordering specific statements. 
Alg.~\ref{alg:patch} shows the patch generation algorithm of ContractFix.

\subsubsection{Program Analysis Infrastructure}
\label{subsubsec:programanalysis}

ContractFix customized static program analysis techniques to extract the necessary context information for vulnerabilities in smart contracts, including involved variables and their control and data dependencies.
We next describe the static program analysis techniques employed by ContractFix.
\input{tables/fixpattern.tex}
\input{patchAlg.tex}

\textbf{Intra-Procedural Data-flow Analysis} 
ContractFix performs an intra-procedural data-flow analysis to collect the program points (i.e. statements) where a variable is created, read, modified, and deleted~\cite{compiler,programanalysis}.
Our intra-procedural data-flow analysis starts with building the method's control flow graph (CFG), where each statement is considered as a single basic block for the convenience of dependency analysis.
It is worth mentioning that modifiers assigned to methods in smart contracts can be executed both before and after the execution of the method body and the parameters of methods can be used in the modifiers.
Thus, the control flow of a method follows this sequence: modifier, method body, modifier. 
Once the CFG is built, existing data-flow analysis is employed to build the data-flow graph (DFG) for the method.
Note that fixing \textit{Reentrancy} requires inter-procedural analysis, which is achieved by combining the method summaries with the intra-procedural analysis.
\begin{figure}
    \centering
    \includegraphics[width=0.45\textwidth]{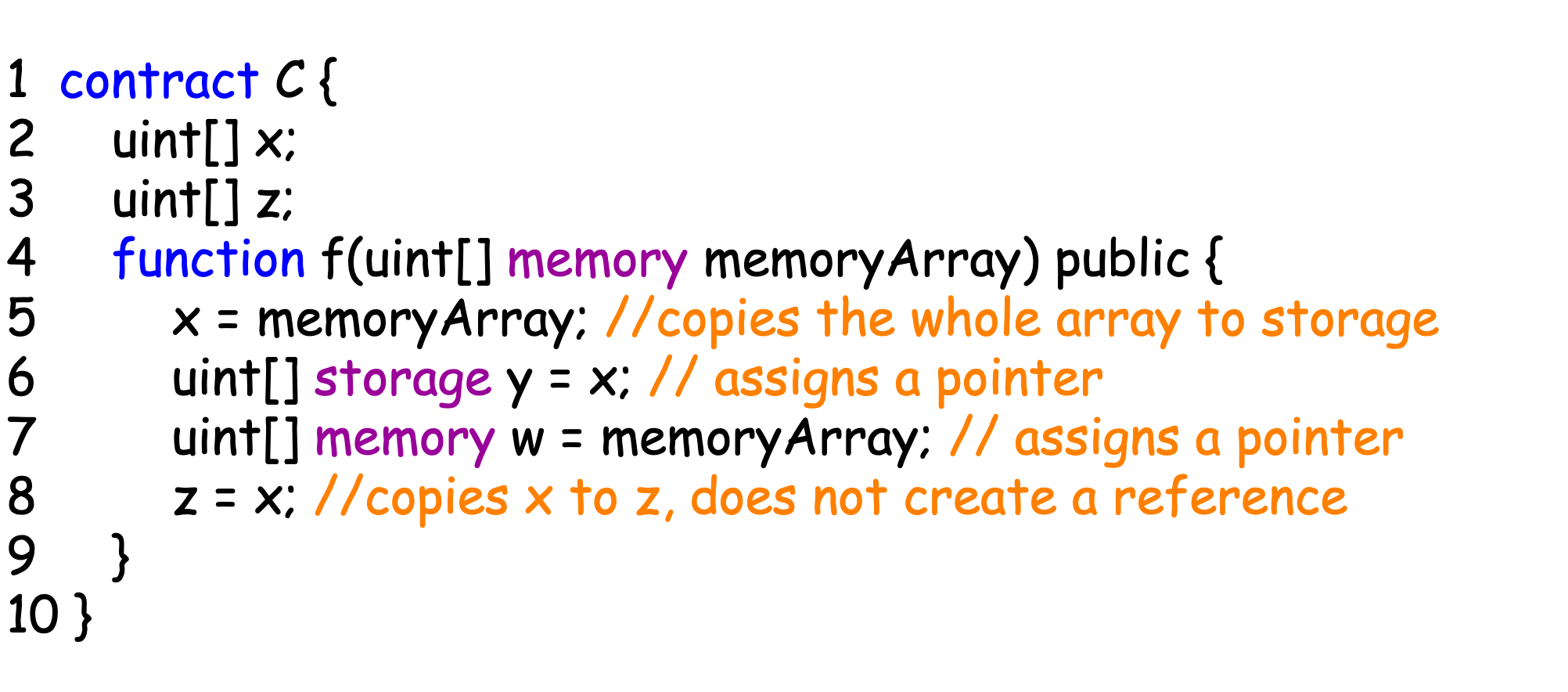}
    \vspace{-10pt}
    \caption{Reference creations in Solidity}
    \label{fig:reference}
\end{figure}

\textbf{Pointer Analysis for Solidity} 
Pointer analysis is known to be expensive and is required for precise analysis (e.g. flow-sensitivity analysis and context-sensitivity analysis).
As smart contract languages such as Solidity restrain the usage of references, existing pointer analysis can be easily adapted for obtaining accurate point-to information for the contracts.
In Solidity ($\geq$v0.6.1), there are three locations where a variable can be stored:
\begin{itemize}
    \item \textit{memory}: the variable in memory is not persistent and its lifetime is limited to an external method call.
    \item \textit{storage}: the variable in storage is persistent and its lifetime is the same as the contract's. 
    \item \textit{calldata}: this location is only available for external method call parameters.
\end{itemize}

Solidity's reference types include $struct$, $array$, and $mapping$. 
Fig.~\ref{fig:reference} shows an example contract to illustrate reference creations.
There are only two situations where a variable of these types can be a reference: 
(1) assignments from a variable in storage to a local variable in storage create a reference (Line 6);
(2) assignments from a variable in memory to another variable in memory create a reference (Line 7).

To determine which pointer analysis algorithm to use, we analyzed $6,420$ real-world smart contracts to find how often the reference type is used in solidity programs. 
We scanned all assignments among these contracts and checked if there are variables that meet the definitions of reference types mentioned above. 
In summary, we found $3,210$ reference variables among $199,724$ assignments in $6,420$ contracts.
That is, on average, only $1.6\%$ of the assignments use reference types.

Thus, ContractFix adapts a flow-insensitive and context-insensitive pointer analysis by extending its point-to model to incorporate data locations and reference creations as described above, and computes the point-to information for each variable in a contract~\cite{compiler}. 

\textbf{Inter-Procedural Analysis via Method Summary} 

To enable inter-procedural analysis, for each method, ContractFix builds a method summary that computes the side effects of the state variables, i.e. whether the state variables are modified in each method. 
Method summaries (or called function summaries) have been used to build inter-procedural program analysis in a modular way~\cite{functionsummary,xiao12:infoflow,functionsummary2},
which can be easily combined with other intra-procedural program analysis to enable inter-procedural analysis.

\textbf{Data Dependency Classification}

Based on the inter-procedural analysis,  ContractFix further classifies data dependency into the following types:
\begin{itemize}
    \item \textit{Flow Dependence} or \textit{Read After Write (RAW):} a statement $s_2$ is flow dependent on $s_1$ if and only if $s_1$ modifies a resource that $s_2$ reads and $s_1$ precedes $s_2$ in execution. 
    \item \textit{Anti-Dependence} or \textit{Write After Read (WAR):} a statement $s_2$ is antidependent on $s_1$ if and only if $s_2$ modifies a resource that $s_1$ reads and $s_1$ precedes $s_2$ in execution.
    \item \textit{Output Dependence} or \textit{Write After Write (WAW):} a statement $s_2$ is output dependent on $s_1$ if and only if $s_1$ and $s_2$ modify the same resource and $s_1$ precedes $s_2$ in execution.
    \item \textit{Input Dependence} or \textit{Read After Read (RAR):} a statement $s_2$ is input dependent on $s_1$ if and only if $s_1$ and $s_2$ read the same resource and $s_1$ precedes $s_2$ in execution.
\end{itemize}

The classification of the data dependency will later be used by the fix patterns (e.g. fixing \textit{Reentrancy} in Sec.~\ref{subsec:methodlevel}) to guide the patch generation.
\subsubsection{Fix Patterns in Statement Level}
Vulnerabilities at this level are usually caused by misuse of individual statements,
such as \textit{UnhandledException} that forgets to check method return values.
\textbf{Fixing \textit{UnhandledException}} 
The fix pattern (Lines 1-6 in Alg.~\ref{alg:patch}) for this vulnerability is to check the return value of each coin transfer function: \textit{send()} and \textit{value()}.
The type of their return values is boolean because they indicate whether the transfers succeed. 
Transactions in which these transfers fail must be reverted to notice the caller to ensure the coherence between the contract states and the transactions. 
Thus, to fix this vulnerability, ContractFix adds a \textit{require()} function call to validate the return values of \textit{send()} and \textit{value()} (Line 4 in Alg.~\ref{alg:patch}) and ensures that their executions are successful before completing the whole transaction. 

\subsubsection{Fix Patterns in Method Level}
\label{subsec:methodlevel}
Vulnerabilities in this level are usually caused by missing parameter checks or miuse some method calls in a method. 
The two types of vulnerabilities in this level are \textit{Reentrancy} and \textit{MissingInputValidation}. 
\textbf{Fixing \textit{Reentrancy}} 
The preferred fix pattern for \textit{Reentrancy} (Lines 8-17 in Alg.~\ref{alg:patch}) is to move all writes to storage ahead so that there is no write to storage after an external method call or a coin transfer call, such as the patch shown in Fig.~\ref{fig:reentrancyExampleDUP}.
ContractFix first identifies the method that has the vulnerability (Lines 9-10), 
and computes the method summary, the pointer information, and the DFG of the method
(Lines 11-13).
ContractFix then identifies the writes that result in the vulnerability (Line 14),
and further computes the data dependencies that are used to move the writes (Line 15).
In particular, if any of these writes (represented as $w$) has data dependencies to the variables used by the external calls (represented as $c$), depending on the type of the dependencies, ContractFix may eliminate such dependencies without changing the semantics before moving the writes ahead:

\begin{itemize}
    \item For \textit{flow dependence} from $w$ to a statement $s$, ContractFix creates a temporary variable to store the value of the variables in $w$ before they are written and replace the same variables in $c$ with these temporary variables, so that $w$ does not impact $c$ if $w$ is moved ahead. 
    \item For \textit{anti-dependence} and \textit{output dependence} from $w$ to a statement $s$, ContractFix moves both $w$ and $s$ ahead if $s$ is not the external call.
    \item For \textit{input dependence} from $w$ to a statement $s$, ContractFix simply moves $w$ ahead since there are no side effects in this type of dependency.
\end{itemize}

However, when there is a \textit{anti-dependence} or \textit{output dependence} from $w$ to $c$, the data dependencies cannot be eliminated. 
Because in this case the updates in $w$ must wait for the execution results of $c$ so $w$ cannot be moved ahead of $c$.
In this case, ContractFix adopts another more expensive fix pattern, which declares a new global \textit{bool} value as a lock to limit the method invocations~\cite{nguyen2021sguard}. 
This lock will not allow the unexpected recall, if the previous call does not finish the execution. 
As the modification of a global variable is much more expensive than the declarations of local variables in smart contracts, we find out that for each transaction, the global lock increases the gas cost by $\sim25000$ for the function, while declaring temporary variables and moving statements increase the gas cost by only $5$. 

\textbf{Fixing \textit{MissingInputValidation}} 
The fix pattern for \textit{MissingInputValidation} is to add conditional checks to validate the method parameters at the beginning of method body (Lines 18-25 in Alg.~\ref{alg:patch}). 
To patch this vulnerability, ContractFix first identifies the 
method parameters that are not validated (Line 21 in Alg.~\ref{alg:patch}).
To do so, ContractFix checks whether the parameters appear in any \textit{require()} method, which is often used in Solidity to perform validation. 
This checking can be easily done by using the DFG of the method. 
ContractFix inserts validations for the other unchecked parameters:

\begin{itemize}
    \item For those parameters whose type is address, ContractFix adds a common validation to check whether this address is $0x0$ because an address with value \textit{0x0} is invalid. 
    \item For parameters whose type are integer, ContractFix adds a safe math library to prevent integer overflow and underflow when doing calculations.
    \item For parameters whose type are bytes or self-defined, ContractFix may not add proper validations because it lacks contextual information for ContractFix to acquire sufficient information about their valid ranges.
\end{itemize}

\begin{figure}[!htb]
    \centering
    \includegraphics[width=0.45\textwidth]{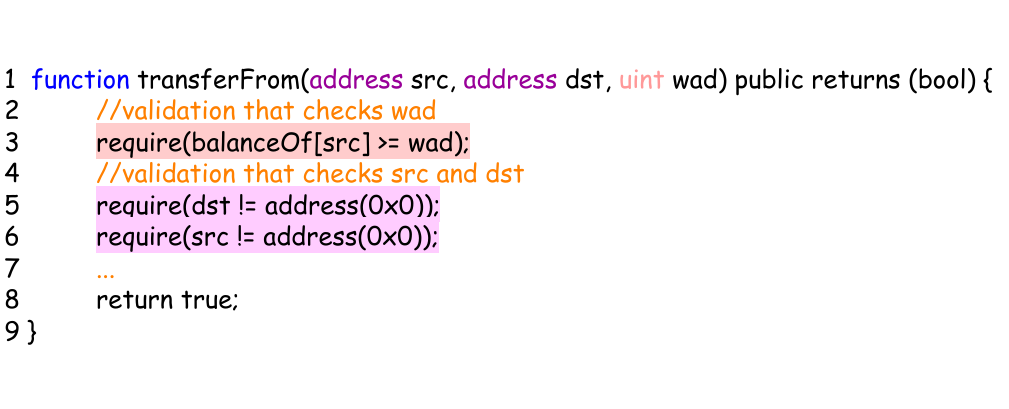}
    \caption{Patch for \textit{MissingInputValidation} vulnerability}
    \label{fig:missinput2}
\end{figure}

Consider the contract in Fig.~\ref{fig:missinput2}. 
The parameter {\color[HTML]{FFCCCC}{\textit{wad}}} is validated by the highlighted statement (Line 6) and thus there is no need for further validation. 
For the parameters {\color[HTML]{990099}{\textit{src}}} and {\color[HTML]{990099}{\textit{dst}}}, ContractFix adds a \textit{require()} method to validate their values (Lines 8-9).

When all the identified vulnerabilities are patched, ContractFix converts its transformed AST back to source code and outputs the patched contract to Phase III.

\subsection{Phase III: Patch Verification}
In this phase, ContractFix reapplies the static verification tools to ensure that the detected vulnerabilities are eliminated while the expected behaviors are preserved in the patched contract.
Once a patched contract candidate is generated, ContractFix applies static verification again on the patched contract and checks whether the vulnerabilities reported in Phase I are eliminated. 
If the static verification tool no longer reports the same vulnerabilities, the patched contract is considered to pass the static verification. 

\section{Evaluation}
\input{tables/eachdetectorres}

We implemented ContractFix in JavaScript ($\sim$5000 lines of code).
We adopted three state-of-the-art vulnerability tools, Securify~\cite{securify}, \textit{Slither}~\cite{feist2019slither}, and \textit{Smartcheck}~\cite{smartcheck}, as the static verification tools for ContractFix.

We built the parsing and transformation modules upon an open source Solidity parser~\cite{solidityparsing} built on top of ANTLR4~\cite{antlr}.
We evaluated the effectiveness of ContractFix in fixing vulnerabilities in real-world smart contracts. 
Specifically, our evaluations aim to answer the following research questions.

\begin{itemize}
    \item \textbf{RQ1:} How effective is ContractFix in generating successful patches for vulnerable contracts?
    \item \textbf{RQ2:} How effective is the synergy of static verification and post-processing in detecting vulnerabilities, compared to static verification only?
    \item \textbf{RQ3:} How efficient is ContractFix in generating patches? 
\end{itemize}
\input{tables/rq1.tex}

\subsection{Evaluation Setup}
Our evaluation datasets have $144$ contracts ($141,404$ lines of Solidity code), and the evaluations are conducted on a server with Intel(R) Xeon(R) CPU E5-2637 v4 (3.50GHz) and 256GB RAM.

\textbf{Injected Vulnerabilities}.
We use a widely adopted dataset~\cite{ghaleb2020effective} that contains smart contracts with various types of injected known vulnerabilities (e.g. integer overflow/underflow, reentrancy, and timestamp-dependency).
We choose $50$ contracts with injected reentrancy vulnerabilities as other types of vulnerabilities in the dataset are not the focus of ContractFix. 
These contracts provide the detailed information on the injected vulnerabilities and thus we can directly evaluate our patch generation without applying vulnerability detection. 
\textbf{Real Vulnerabilities}.
We collect real smart contracts based on the addresses obtained from BigQuery~\cite{Bigquery},
and select the top $10,000$ contracts sorted by the number of transactions.
We further download the source code from Etherscan~\cite{Etherscan} based on the addresses, and exclude the contracts whose source code is not available. 
We then apply static verification on these downloaded contracts and exclude the ones that cannot be analyzed by the Solidity compiler due to version incompatibility.
In total, we obtain $4,640$ contracts.
For each contract to be used in our evaluation, we inspect the source code to confirm vulnerabilities, verify the patches, and prepare test cases that exercise the vulnerable behaviors of the original contracts, which requires non-trivial manual efforts.
Within our affordable efforts, we select $94$ vulnerable contracts that have the four types of vulnerabilities described in Sec.~\ref{subsec:study}, as shown in Tab.~\ref{tab:rq1}.
The test cases are shown in Tab.~\ref{tab:rq2patch}.
If a contract has several types of vulnerabilities, we classify it into the \textit{Mixed} category.

\textbf{Semantic Validation}.
To ensure that the patched contracts preserve the expected behaviors, semantic validation is conducted using a smart contract testing platform named Truffle~\cite{tuffle}. 
Truffle allows us to deploy a contract on a local blockchain powered by Ganache and make a method call or issue a transaction. 
To obtain the test cases for checking contracts' expected behaviors, we made use of the publicity of all transactions on blockchain: we downloaded the existing transactions of a contract and extracted the input values to create test cases because these transactions should reflect the functionality of a contract. 
But these constructed test cases lack test oracles to assert the expected results. 
To address this problem, we adopt the idea of multiple implementation testing~\cite{multitest1,testingbook}: we consider the original contract and the patched contract as different implementations for the same requirements and assert that the states of both contracts should be the same after the testing. 
Specifically, ContractFix first deploys both the patched contract and the original contract on the Truffle with the same initial states. 
Then, ContractFix runs the test cases and compares the states of both contracts. 
If the states of both contracts remain the same after the testing, ContractFix considers that the expected behaviors are preserved in the patched contract.

\subsection{RQ1: Effectiveness in Patch Generation}
\label{subsec:eval-rq2}

\input{tables/rq2.tex}

We run ContractFix on each of the contracts in our testing set to generate a patched contract.
We then manually examine the patched contract and verify its correctness. 
We also examine why ContractFix fails to generate patches for certain contracts.

\textbf{Overall Results} 
Tab.~\ref{tab:rq2patch} shows the path generation results.
On average, each contract in the testing set has $4.17$ vulnerabilities.
We consider a patch is successful if the vulnerability to fix cannot be detected by the static verification tools in the patched contract and the patched contract passes the corresponding test cases, including the test cases written to exercise the vulnerable behaviors (Tab.~\ref{tab:rq2patch}) and the test cases built from the public-transaction records (Tab.~\ref{tab:replay}).
For contracts with injected vulnerabilities, we perform manual verification since these contracts lack public-transaction records for writing test cases. 
Overall, ContractFix successfully generates $565$ patches for $601$ vulnerabilities, achieving a very high success rate ($94\%$).
These results indicate that the combination of template-based fix patterns and static program analysis is very effective in generating successful patches.

\textbf{Patch Statistics}
For the real contracts, on average, ContractFix needs to change $3.7$, $17.47$, $3.71$, $2.26$, and $12.71$ lines of code to patch a contract with \textit{Reentrancy} vulnerabilities, \textit{LockedEther} vulnerabilities, \textit{MissingInputValidation} vulnerabilities, \textit{UnhandledException} vulnerabilities, and multiple types of vulnerabilities (i.e.\textit{Mixed}). 
Without considering the type of vulnerabilities, ContractFix needs to averagely change $7.97$ lines of code to patch a contract. 
For the contracts with injected vulnerabilities, ContractFix needs to change $4$ lines of code to patch the \textit{Reentrancy} vulnerabilities.

\textbf{Transaction Replay}
To show that the patch preserves the original functionality of the smart contracts, we chose 25 contracts (top 5 contracts that have the most transactions in each vulnerable type) and crawled $5000$ transactions for each of them ($125,000$ transactions in total).
We then deployed their original version and the patched version in the local testing environment (e.g. Ganache), and issued transactions to execute the contracts using the crawled transaction data. 
Tab.~\ref{tab:replay} shows the results of the transaction replay. 
Column ``Status diff'' shows the number of different result states.
Column ``Gas Diff'' shows the average gas usage differences.
We can see the number of contracts that have different result states are all $0$. 
This result shows that \textit{the patches generated by ContractFix preserves the contracts' original functionality}.
The results also show that gas usage has only a slight increase (maximum being $188.84$).
As each unit of gas equals to $10E-9$ Ethers~\cite{gasprice}, which is $\sim\$0.000027$, \textit{the cost of the extra gas is negligible}.

\textbf{Failed Patches}
For real contacts, ContractFix fails to generate patches for $2$ \textit{Reentrancy} vulnerabilities,
which is mainly due to the stack depth limit of EVM. 
When fixing \textit{Reentrancy} vulnerability, ContractFix usually needs to create temporary variables. 
While not often, such behavior may trigger the EVM exceptions about stack depth, causing the patches not to pass the validation. 
This case may be improved by requesting the developers to limit their call stack, which will also help defend stack depth attack~\cite{stackdepth}. 
The main reason why fixing \textit{MissingInputValidation} and \textit{LockedEther} may fail lies in the limitation of our source code generator,
which adopts the pre-order traversal of the patched AST to generate patches.
However, when a method call is used as the argument for a function modifier, 
the generated source code will have syntax errors.
This case can be fixed by improving the code generation mechanism.

For contracts with injected vulnerabilities, ContractFix fails to generate patches for $30$ contracts as our dataflow analysis considers the injected vulnerabilities cannot be fixed by moving statements. 
But \textit{all these $30$ contracts can be fixed by ContractFix by adopting the global
bool value as a lock to limit the method invocations}.

\textbf{Comparison to Existing Works}.
As there are no existing works that combine detection, patching, and verification as ContractFix does, we cannot directly compare ContractFix with existing works.  
Thus, we compare only the patching step in our evaluations. 
As shown in Tab.~\ref{tab:patterns}, ContractFix provides fix patterns (global lock for \textit{Reentrancy} and owner check for \textit{LockedEther}) that EVMPatch~\cite{rodler2021evmpatch} and sGuard~\cite{nguyen2021sguard} can support. 
The results show that ContractFix can effectively generate these patches (91\% for \textit{Reentrancy} and 99\% for \textit{LockedEther}) as the existing works without compromising expected behaviors. 
Beyond that, ContractFix supports moving statement to fix \textit{Reentrancy} vulnerabilities with a much lower gas cost than sGuard (5 v.s. 25000).
EVMPatch cannot this fix pattern as binary code loses the source code semantics and requires extra data analysis to move the statements.
\begin{lstlisting}[numbers = none]
+ var totalUnreleasedTokens_temp = totalUnreleasedTokens.
+ vestingSchedule.principleLockAmount = _principleLockAmount;
+ vestingSchedule.bonusLockAmount = _bonusLockAmount;
+ vestingSchedule.isPrincipleReleased = false;
+ vestingSchedule.isBonusReleased = false;
+ totalUnreleasedTokens = safeAdd(totalUnreleasedTokens,         _totalAmount);
+ vestingSchedule.amountReleased = 0;
+ require(token.balanceOf(this) >= safeAdd(totalUnreleasedTokens_temp, _totalAmount));

- require(token.balanceOf(this) >=                               safeAdd(totalUnreleasedTokens, _totalAmount));
- vestingSchedule.principleLockAmount = _principleLockAmount;
- vestingSchedule.bonusLockAmount = _bonusLockAmount;
- vestingSchedule.isPrincipleReleased = false;
- vestingSchedule.isBonusReleased = false;
- totalUnreleasedTokens = safeAdd(totalUnreleasedTokens,         _totalAmount);
- vestingSchedule.amountReleased = 0;
\end{lstlisting}
Based on the data-flow analysis, ContractFix finds that there is a WAR dependence between \textit{safeAdd()} and the writes to \textit{totalUnreleasedTokens}.
In this case, ContractFix generates a patch that saves \textit{totalUnreleasedTokens} before \textit{safeAdd()} by creating a temporary variable, replaces the parameters of \textit{safeAdd()} with the temporary variable, and moves all the writes ahead.

\begin{table}[t]
\centering
\caption{Execution results about transaction replay}
\label{tab:replay}
\begin{tabular}{crr}
\hline
\textbf{Type}       & \multicolumn{1}{c}{\textbf{Status Diff.}} & \multicolumn{1}{c}{\textbf{Gas Diff.}} \\ \hline
\textbf{Reentrancy}          & 0                                   & 53.38                      \\
\textbf{MissinputValidation} & 0                                   & 46.99                        \\
\textbf{LockedEther}         & 0                                   & 7.36                           \\
\textbf{UnhandledException}  & 0                                   & 27.68                      \\
\textbf{Mixed}               & 0                                   & 188.84                      \\ \hline
\end{tabular}
\end{table}

\subsection{RQ2: Ensemble of Static Verification Tools}
\label{subsec:eval-rq1}

In this RQ, we evaluate the effectiveness of ContractFix's ensemble of multiple static verification tools. 
Tab~\ref{tab:detectors} shows the vulnerability detected by different static verification tools by ContractFix.

Column \textit{Majority} shows the number of vulnerabilities confirmed by at least two static verification tools except for \textit{MissInputValidation}, because only \textit{Securify} supports the detection of this vulnerability.
The results show that our post-processing filters out the candidates that cannot be fixed (Sec.~\ref{subsec:detection}).
For example, \textit{MissingInputValidation} reports 795 vulnerabilities and only 131 of them are related to address types. 
We further manually examine the detected vulnerabilities by the majority voting, and confirm all of them are true positives, indicating the effectiveness of majority voting in improving the detection performance.
For example, \textit{Slither} reports 129 \textit{UnhandledException} vulnerabilities, while the majority voting confirms 60 out of them.
Similarly, \textit{Slither} misses 37 \textit{LockedEther} vulnerabilities, but the combination of \textit{Securify} and \textit{SmartCheck} finds these 37 vulnerabilities.

We can also see the combination of post-processing and majority voting addresses the limitations of using only one static verification tool.
For example, some security properties used by \textit{Securify} are too general: for \textit{Reentrancy} vulnerabilities, \textit{Securify}'s property detects all the writes to storage after an external method call; however, if another external method call is used to determine the execution of the writes to storage, a false positive is reported.
Based on the results, \textit{Securify} reports 26 false positives, which is first reduced by the post-processing to 3 and then by the majority voting to 0.
These results demonstrate that static-verification and post-processing greatly improve the precision of the vulnerability detection, making it feasible and practical to support the patch generation.

\subsection{RQ3: Runtime Performance}
\label{subsec:eval-rq3}

To understand the performance of ContractFix, we measure the execution time of ContractFix's three phases. 
Tab.~\ref{tab:rq3time} shows the results. 
Column \textit{Detection} shows the execution time for pre-processing.
Column \textit{Patch} shows the execution time for patch generation. 
Column \textit{Verification} shows the execution time for patch verification using Securify. 
We exclude the validation using Truffle since it requires manual interactions (e.g. sending transactions).
As we can see, ContractFix takes $1159.58s$ to finish the whole process. 
Without considering the time needed by static verification (i.e. detection and validation), ContractFix only takes $3.75s$ to fix a contract on average, indicating that ContractFix's light-weight program analysis and patch generation based on template-based fix patterns are very efficient.

\input{tables/rq3.tex}

\section{Discussion}
\label{sec:discuss}

\textbf{Static Verification}
Static verification unavoidably produces false positives.
A major reason is that some security properties are too general and cannot describe various specific behaviors (Sec.~\ref{subsec:detection}).
ContractFix addresses this problem by leveraging majority voting to ensemble multiple static verification tools and employs post-processing to filter out  vulnerabilities not amenable for automatic fix.
Note that the patch generation of ContractFix does not depend on the intermediate information of any detectors, and thus can be integrated  with various types of detectors.
Post-processing is relatively easy to extend as it focuses on only local context (i.e. mainly intra-procedural analysis), while static verification considers the global context and is difficult to customize.
Alternatively, more precise static verification with more flexible security properties can be used, but this direction requires further research efforts and is out of the scope of this paper.

\textbf{Generalization of Fix Patterns}
Similar to Solidity, most smart contract languages restrain the uses of references in the language level, such as Move~\cite{move},
and ContractFix can be extended to support the fix patterns of these languages.

\textbf{Limitations}
ContractFix cannot fix vulnerabilities whose exploits rely on the mechanisms of the underlying blockchain platform, 
such as transaction-executing order (e.g. \textit{TODAmount} and \textit{TODReceiver} vulnerabilities) and timestamp.
Existing research on blockchain security~\cite{karame2016bitcoin,ekiden,solidus,kosba2016hawk} can be adopted to address these vulnerabilities, being beyond the scope of this paper. 

\section{Related Work}
\label{sec:literature}

\textbf{Smart Contract Vulnerabilities}
The openness and the immutability of smart contract execution introduce significant vulnerabilities against various attacks. 
Early work~\cite{luu2016making,zeus,maian} categorizes contract vulnerabilities and detects them using symbolic execution, model checking, and trace analysis. 
Recent work~\cite{securify,callback} refines the checking using specific patterns and develops verification techniques to detect pattern violations as vulnerabilities.
Rodler { et.al.}~\cite{rodler2018sereum} further propose to detect reentrancy attacks using dynamic taint tracking.
ContractFix leverages this line of research to detect contract vulnerabilities and verify removal of them.

\textbf{Automatic Program Repair} 
Some previous  research~\cite{apr,genprog,aprcontext,testrepair,AutoBinary} focuses on the code that is executed for negative test cases,  and then produces modifications to a program, 
including deleting a statement and inserting a statement found in the program.
Treating all the modifications of a program as a search space, the research  adapts search algorithms such as genetic algorithms to generate patches,  facing significant challenges in search space explosion~\cite{aprchallenges,aprcontext}.
Some other research~\cite{nguyen2013semfix,mechtaev2016angelix} uses constraint solving to find correct expressions to replace incorrect or vulnerable expressions in a program. 
Compared with the preceding previous research, ContractFix does not rely on the test suite that triggers a failure of the target program to generate a  patch. 
Also, ContractFix conducts static verification and rule-based checking to detect vulnerabilities and includes program analysis techniques to enable complex fix patterns.

\section{Conclusion}
In this paper, we have presented a novel framework named ContractFix, which automatically generates source code patches to fix
vulnerabilities in a smart contract.
ContractFix is powered by three key designs:
(1) ContractFix synergistically combines three static verification tools and conducts post-processing to detect vulnerabilities for a smart contract without a need for test cases;
(2) ContractFix includes program analysis techniques tailored to smart contracts to enable fix patterns to include complex repair actions such as introduction of temporary variables;
(3) ContractFix conducts static verification to prove the elimination of the vulnerabilities in the patched contract.
Our evaluations on vulnerable contracts 
from a widely used dataset and real-world vulnerable contracts have shown that ContractFix can successfully fix $94\%$ of the detected vulnerabilities.

\bibliographystyle{unsrt}  
\bibliography{references}

\end{document}

%% file: tables/fixpattern.tex
\begin{table}[t]
\centering
\caption{Summary of fix patterns}
\label{tab:patterns}
\resizebox{0.49\textwidth}{!}{
\begin{tabular}{ccr}
\hline
\textbf{Type}          & \multicolumn{1}{c}{\textbf{Level}}          & \multicolumn{1}{c}{\textbf{Fix Pattern}} \\ \hline
\textbf{Reentrancy}                      &   Method           & Lock~\cite{reentrybug,nguyen2021sguard}, Reorder statements~\cite{reentrybug}        \\
\textbf{MissingInputValidation}         &    Method           & Require check~\cite{missingzerocheck}                   \\
\textbf{LockedEther}                      &    Contract         & Withdraw function~\cite{addwithdraw}  \\ 
\textbf{UnhandledException}              &   Statement           & Require check~\cite{swc104}                   \\ \hline
              
\end{tabular}
}
\end{table}

%% file: patchAlg.tex
\begin{algorithm}[!ht]
    \caption{Security Patch Generation of ContractFix}\label{alg:patch} 
    \begin{algorithmic}[1]
    \REQUIRE $ast$ as the AST of original source code, 
    $violations$ as the detected vulnerabilities
    \ENSURE $p\_ast$ as the patched AST
    \STATE $//$ Patching UnhandledException
    \FOR{each $vulnerability$ in $violations.UnhandledException$}
    \STATE $call \leftarrow vulnerability.line$
    \STATE $validations \leftarrow$ addCheck$(call)$
    \STATE $patch \leftarrow$ patchGenerator$(validations)$
    \STATE $p\_ast \leftarrow$ ASTTransformer$(patch, ast)$
    \ENDFOR
    \STATE $//$ Patching Reentrancy
    \FOR{each $vulnerability$ in $violations.Reentrancy$}
    \STATE $call \leftarrow vulnerability.line$
    \STATE $func \leftarrow$ locateFunc$(ast, call)$
    \STATE $func\_sums \leftarrow$ scan$(ast)$
    \STATE $pointer\_info \leftarrow$ pointerAnalysis$(ast)$
    \STATE $DFG \leftarrow$ dataflowAnalysis$(ast, func,$ \\ \qquad \qquad $pointer\_info, func\_sums)$
    \STATE $writes \leftarrow$ findWritestoStorage$(func, call)$
    \STATE $dependences \leftarrow$ dependencyAnalysis$(DFG, writes,$ \\ $ call, func)$
    \STATE $patch \leftarrow$ patchGenerator$(dependences, func)$
    \STATE $p\_ast \leftarrow$ ASTTransformer$(patch, ast)$
    \ENDFOR
    \STATE $//$ Patching MissingInputValidation
    \FOR{each $vulnerability$ in $violations.MissingInputValidation$}
    \STATE $func \leftarrow$ locateFunc$(vulnerability.line)$
    \STATE $unchecked\_parameters \leftarrow$ checkValidation$(func.arguments)$
    \STATE $validations \leftarrow$ addValidation$(unchecked\_parameters)$
    \STATE $patch \leftarrow$ patchGenerator$(validations)$
    \STATE $p\_ast \leftarrow$ ASTTransformer$(patch, ast)$
    \ENDFOR
    \STATE $//$ Patching LockedEther
    \FOR{each $vulnerability$ in $violations.LockedEther$}
    \STATE $contract \leftarrow$ locateContract$(vulnerability.line)$
    \IF{mustPatch$(contract) == True$}
    \STATE $owner \leftarrow$ findOwner$(contract)$
    \STATE $withdraw \leftarrow$ createWithdraw$(owner)$
    \STATE $patch \leftarrow$ patchGenerator$(withdraw)$ 
    \STATE $p\_ast \leftarrow$ ASTTransformer$(patch, ast)$
    \ENDIF
    \ENDFOR
    \end{algorithmic}
\end{algorithm}

%% file: tables/eachdetectorres.tex
\begin{table}[t]
\centering
\caption{Vulnerabilities detected by each detector}
\label{tab:detectors}
\resizebox{0.49\textwidth}{!}{
\begin{tabular}{crrrr}
\hline
\textbf{Type}                   & \multicolumn{1}{c}{\textbf{Securify}} & \multicolumn{1}{c}{\textbf{Slither}} & \multicolumn{1}{c}{\textbf{Smartcheck}} & \textbf{Majority} \\ \hline
\textbf{Reentrancy}             & 107                                   & 461                                  & 0                                       & 23                \\

\textbf{MissingInputValidation} & 979                                   & 0                                    & 0                                       & 131               \\
\textbf{LockedEther}            & 184                                   & 100                                  & 43                                      & 137               \\
\textbf{UnhandledException}     & 83                                    & 129                                  & 36                                      & 60                \\

\textbf{Total}                  & 1353                  & 690                 & 79                    & 351               \\ \hline
\end{tabular}
}
\end{table}

%% file: tables/rq1.tex
\begin{table}[t]
\centering
\caption{Vulnerable contracts used in evaluation}
\label{tab:rq1}
\resizebox{0.45\textwidth}{!}{
\begin{tabular}{crr}
\hline
\textbf{Type}                   & \multicolumn{1}{c}{\textbf{Contract Count}} & \multicolumn{1}{c}{\textbf{Lines of Code}} \\ \hline
\textbf{Reentrancy}                    & 17                           & 8,522                    \\
\textbf{MissingInputValidation} & 21                           & 26,102                   \\
\textbf{LockedEther}           & 20                           & 17,277                   \\
\textbf{UnhandledException}    & 19                           & 30,349                   \\
\textbf{Mixed}                  & 17                           & 38,644                   \\
\textbf{Injected}              &50                             & 20,510 \\
\textbf{Total}         & 144                           & 141,404                  \\ \hline
\end{tabular}
}
\end{table}

%% file: tables/rq2.tex
\begin{table}[t]
\centering
\caption{Effectiveness of ContractFix in patch generation}
\label{tab:rq2patch}
\resizebox{0.48\textwidth}{!}{
\begin{tabular}{crrrrr}
\hline
\textbf{Type}                   & \multicolumn{1}{c}{\textbf{Total}} & \multicolumn{1}{c}{\textbf{Success}} & \multicolumn{1}{c}{\textbf{Fail}} & \multicolumn{1}{c}{\textbf{Suc. Rate}} & \multicolumn{1}{c}{\textbf{Test Case}} \\ \hline
\textbf{Reentrancy}             & 23                                 & 21                                   & 2                                 & 0.91                                   & 73                                       \\
\textbf{MissingInputValidation} & 131                                & 128                                  & 3                                 & 0.98                                   & 68                                       \\
\textbf{LockedEther}            & 137                                & 136                                  & 1                                 & 0.99                                   & 60                                       \\
\textbf{UnhandledException}     & 60                                 & 60                                   & 0                                 & 1.00                                   & 245                                      \\
\textbf{Injected}               & 250     & 220 & $^*$30 & 0.88 &  -\\
\textbf{Total}                  & 601                                & 565                                  & 36                                 & 0.94                                   & 446                                      \\ \hline
\end{tabular}
}
\end{table}

%% file: tables/rq3.tex
\begin{table}[t]
\centering
\caption{Runtime performance of ContractFix}
\label{tab:rq3time}
\resizebox{0.45\textwidth}{!}{
\begin{tabular}{crrr}
\hline
\textbf{Type}                   & \multicolumn{1}{c}{\textbf{Detection (s)}} & \multicolumn{1}{c}{\textbf{Patch (s)}} & \multicolumn{1}{c}{\textbf{Validation (s)}} \\ \hline
\textbf{Reentrancy}                    & 641.61                     & 12.56                      & 595.58                      \\
\textbf{MissingInputValidation} & 765.61                     & 0.44                     & 468.17                      \\
\textbf{LockedEther}            & 781.75                     & 0.82                      & 802.73                      \\

\textbf{UnhandledException}     & 84.07                      & 0.33                     & 108.28                      \\
\textbf{Mixed}                  & 855.05                     & 4.62                      & 871.44                     \\
\textbf{AVG}           & 586.59                    & 3.75                    & 569.24                           \\ \hline
\end{tabular}
}
\end{table}

%% file: templateArxiv.bbl
\begin{thebibliography}{10}

\bibitem{nakamotobitcoin}
Satoshi Nakamoto.
\newblock Bitcoin: A peer-to-peer electronic cash system.

\bibitem{smartcontract}
Vitalik Buterin.
\newblock Ethereum: a next generation smart contract and decentral- ized
  application platform, 2013.
\newblock https://github.com/ethereum/ wiki/wiki/White-Paper.

\bibitem{blockchain-finance}
Blockchain in financial services, 2021.
\newblock https://consensys.net/blockchain-use-cases/finance/.

\bibitem{blockchain-casino}
Blockchain casino games, 2021.
\newblock https://blockchain-casino-games.com/.

\bibitem{blockchain-identity}
Blockchain in digital identity, 2021.
\newblock https://consensys.net/blockchain-use-cases/digital-identity/.

\bibitem{solidity}
Solidity, 2021.
\newblock https://github.com/ethereum/solidity.

\bibitem{Ethereum}
Ethereum platform, 2021.
\newblock https://ethereum.org/.

\bibitem{King}
King of ether, 2021.
\newblock
  \url{https://github.com/kieranelby/KingOfTheEtherThrone\\/blob/master/contracts/KingOfTheEtherThrone.sol}.

\bibitem{frozen}
Accidental’ bug may have frozen \$280 million worth of\\ digital coin ether
  in a cryptocurrency wallet, 2017.
\newblock
  https://www.cnbc.com/2017/11/08/accidental-bug-may-have-frozen-280-worth-of-ether-on-parity-wallet.html.

\bibitem{read}
How to find \$10m just by reading the blockchain, 2021.
\newblock
  https://medium.com/golem-project/how-to-find-10m-by-just-reading-blockchain-6ae9d39fcd95.

\bibitem{atzei2017survey}
Nicola Atzei, Massimo Bartoletti, and Tiziana Cimoli.
\newblock A survey of attacks on ethereum smart contracts (sok).
\newblock In {\em International conference on principles of security and
  trust}, pages 164--186. Springer, 2017.

\bibitem{dao}
The dao attack, 2021.
\newblock
  \url{https://www2.deloitte.com/ie/en/pages/technology/articles/DAO-Attack-Analysis.html}.

\bibitem{parity}
Security alert, 2017.
\newblock https://www.parity.io/security-alert/.

\bibitem{thedao}
{TheDAO}, 2021.
\newblock
  \url{https://etherscan.io/token/0xbb9bc244d798123fde783fcc1c72d3bb8c189413}.

\bibitem{reentrybug}
Reentrancy, 2021.
\newblock https://swcregistry.io/docs/SWC-107.

\bibitem{luu2016making}
Loi Luu, Duc-Hiep Chu, Hrishi Olickel, Prateek Saxena, and Aquinas Hobor.
\newblock Making smart contracts smarter.
\newblock In {\em Proceedings of the 2016 ACM SIGSAC conference on computer and
  communications security}, pages 254--269, 2016.

\bibitem{mythril}
Mythril, 2021.
\newblock https://github.com/ConsenSys/mythril.

\bibitem{securify}
Petar Tsankov, Andrei Dan, Dana Drachsler-Cohen, Arthur Gervais, Florian
  Buenzli, and Martin Vechev.
\newblock Securify: Practical security analysis of smart contracts.
\newblock In {\em Proceedings of the 2018 ACM SIGSAC Conference on Computer and
  Communications Security}, pages 67--82. ACM, 2018.

\bibitem{maian}
Ivica Nikoli{\'c}, Aashish Kolluri, Ilya Sergey, Prateek Saxena, and Aquinas
  Hobor.
\newblock Finding the greedy, prodigal, and suicidal contracts at scale.
\newblock In {\em Proceedings of the 34th Annual Computer Security Applications
  Conference}, pages 653--663. ACM, 2018.

\bibitem{zeus}
Sukrit Kalra, Seep Goel, Mohan Dhawan, and Subodh Sharma.
\newblock Zeus: Analyzing safety of smart contracts.
\newblock In {\em 25th Annual Network and Distributed System Security
  Symposium, {NDSS} 2018, San Diego, California, USA, February 18-21, 2018},
  2018.

\bibitem{rodler2018sereum}
Michael Rodler, Wenting Li, Ghassan~O Karame, and Lucas Davi.
\newblock Sereum: Protecting existing smart contracts against re-entrancy
  attacks.
\newblock {\em arXiv preprint arXiv:1812.05934}, 2018.

\bibitem{nguyen2021sguard}
Tai~D Nguyen, Long~H Pham, and Jun Sun.
\newblock sguard: Towards fixing vulnerable smart contracts automatically.
\newblock {\em arXiv preprint arXiv:2101.01917}, 2021.

\bibitem{avoidbugs}
Amir Michail and Tao Xie.
\newblock Helping users avoid bugs in gui applications.
\newblock In {\em Proceedings of the International Conference on Software
  Engineering (ICSE)}, page 107–116, 2005.

\bibitem{smartcontract-practice}
Ethereum smart contract best practices, 2021.
\newblock https://consensys.github.io/smart-contract-best-practices/.

\bibitem{le2011genprog}
Claire Le~Goues, ThanhVu Nguyen, Stephanie Forrest, and Westley Weimer.
\newblock Genprog: A generic method for automatic software repair.
\newblock {\em Ieee transactions on software engineering}, 38(1):54--72, 2011.

\bibitem{nguyen2013semfix}
Hoang Duong~Thien Nguyen, Dawei Qi, Abhik Roychoudhury, and Satish Chandra.
\newblock Semfix: Program repair via semantic analysis.
\newblock In {\em 2013 35th International Conference on Software Engineering
  (ICSE)}, pages 772--781. IEEE, 2013.

\bibitem{mechtaev2016angelix}
Sergey Mechtaev, Jooyong Yi, and Abhik Roychoudhury.
\newblock Angelix: Scalable multiline program patch synthesis via symbolic
  analysis.
\newblock In {\em Proceedings of the 38th international conference on software
  engineering}, pages 691--701. ACM, 2016.

\bibitem{long2016automatic}
Fan Long and Martin Rinard.
\newblock Automatic patch generation by learning correct code.
\newblock In {\em ACM SIGPLAN Notices}, volume~51, pages 298--312. ACM, 2016.

\bibitem{testrepair}
Yingfei Xiong, Xinyuan Liu, Muhan Zeng, Lu~Zhang, and Gang Huang.
\newblock Identifying patch correctness in test-based program repair.
\newblock In {\em Proceedings of the International Conference on Software
  Engineering (ICSE)}, page 789–799, 2018.

\bibitem{staticrepair}
Rijnard van Tonder and Claire {Le Goues}.
\newblock Static automated program repair for heap properties.
\newblock In {\em Proceedings of the 40th International Conference on Software
  Engineering, {ICSE} 2018, Gothenburg, Sweden, May 27 - June 03, 2018}, pages
  151--162, 2018.

\bibitem{genprog}
Claire {Le Goues}, Michael Dewey{-}Vogt, Stephanie Forrest, and Westley Weimer.
\newblock A systematic study of automated program repair: Fixing 55 out of 105
  bugs for {\textdollar}8 each.
\newblock In {\em 34th International Conference on Software Engineering, {ICSE}
  2012, June 2-9, 2012, Zurich, Switzerland}, pages 3--13, 2012.

\bibitem{AutoBinary}
Eric Schulte, Jonathan DiLorenzo, Westley Weimer, and Stephanie Forrest.
\newblock Automated repair of binary and assembly programs for cooperating
  embedded devices.
\newblock In {\em Proceedings of the Eighteenth International Conference on
  Architectural Support for Programming Languages and Operating Systems},
  ASPLOS ’13, page 317–328, New York, NY, USA, 2013. Association for
  Computing Machinery.

\bibitem{aprchallenges}
Claire {Le Goues}, Stephanie Forrest, and Westley Weimer.
\newblock Current challenges in automatic software repair.
\newblock {\em Software Quality Journal}, 21(3):421--443, 2013.

\bibitem{aprcontext}
Ming Wen, Junjie Chen, Rongxin Wu, Dan Hao, and Shing-Chi Cheung.
\newblock Context-aware patch generation for better automated program repair.
\newblock In {\em Proceedings of the International Conference on Software
  Engineering (ICSE)}, page 1–11. Association for Computing Machinery, 2018.

\bibitem{feist2019slither}
Josselin Feist, Gustavo Grieco, and Alex Groce.
\newblock Slither: a static analysis framework for smart contracts.
\newblock In {\em 2019 IEEE/ACM 2nd International Workshop on Emerging Trends
  in Software Engineering for Blockchain (WETSEB)}, pages 8--15. IEEE, 2019.

\bibitem{smartcheck}
Sergei Tikhomirov, Ekaterina Voskresenskaya, Ivan Ivanitskiy, Ramil Takhaviev,
  Evgeny Marchenko, and Yaroslav Alexandrov.
\newblock Smartcheck: Static analysis of ethereum smart contracts.
\newblock In {\em Proceedings of the 1st International Workshop on Emerging
  Trends in Software Engineering for Blockchain}, pages 9--16, 2018.

\bibitem{majorityvoting}
Dymitr Ruta and Bogdan Gabrys.
\newblock Classifier selection for majority voting.
\newblock {\em Information fusion}, 6(1):63--81, 2005.

\bibitem{templaterepair}
Kui Liu, Anil Koyuncu, Dongsun Kim, and Tegawend{\'{e}}~F. Bissyand{\'{e}}.
\newblock Tbar: revisiting template-based automated program repair.
\newblock In Dongmei Zhang and Anders M{\o}ller, editors, {\em Proceedings of
  the 28th {ACM} {SIGSOFT} International Symposium on Software Testing and
  Analysis, {ISSTA} 2019, Beijing, China, July 15-19, 2019}, pages 31--42.
  {ACM}, 2019.

\bibitem{ghaleb2020effective}
Asem Ghaleb and Karthik Pattabiraman.
\newblock How effective are smart contract analysis tools? evaluating smart
  contract static analysis tools using bug injection.
\newblock In {\em Proceedings of the 29th ACM SIGSOFT International Symposium
  on Software Testing and Analysis}, 2020.

\bibitem{Etherscan}
Etherscan, 2021.
\newblock https://cn.etherscan.com/.

\bibitem{projectweb}
Contractfix project website, 2022.
\newblock https://github.com/research1132/ContractFix.

\bibitem{bitcoinscript}
Bitcoin script, 2021.
\newblock https://en.bitcoin.it/wiki/Script.

\bibitem{andrychowicz2014secure}
Marcin Andrychowicz, Stefan Dziembowski, Daniel Malinowski, and Lukasz Mazurek.
\newblock Secure multiparty computations on bitcoin.
\newblock In {\em 2014 IEEE Symposium on Security and Privacy}, pages 443--458.
  IEEE, 2014.

\bibitem{hyperledger}
Hyperledger, 2021.
\newblock https://www.hyperledger.org/.

\bibitem{corda}
Corda, 2021.
\newblock https://www.corda.net/.

\bibitem{atzei2016survey}
Nicola Atzei, Massimo Bartoletti, and Tiziana Cimoli.
\newblock A survey of attacks on ethereum smart contracts.
\newblock {\em IACR Cryptology ePrint Archive}, 2016:1007, 2016.

\bibitem{missingzerocheck}
Missing zero check, 2021.
\newblock
  \url{https://github.com/crytic/slither/wiki/Detector-Documentation#missing-zero-address-validation}.

\bibitem{addwithdraw}
Lock ether, 2021.
\newblock
  \url{https://github.com/crytic/slither/wiki/Detector-Documentation#contracts-that-lock-ether}.

\bibitem{swc104}
Swc 104, 2021.
\newblock https://swcregistry.io/docs/SWC-104.

\bibitem{compiler}
Aho Alfred V.; Sethi Ravi ; Ullman~Jeffrey D.
\newblock {\em Compilers: Principles, Techniques, and Tools}.
\newblock 1986.
\newblock {ISBN:} 0-201-10088-6.

\bibitem{programanalysis}
C.~Hankin Nielson~F., H.R.~Nielson;.
\newblock {\em Principles of Program Analysis}.
\newblock 2005.
\newblock {ISBN:} 3-540-65410-0.

\bibitem{functionsummary}
Dongyoon Lee, Peter~M. Chen, Jason Flinn, and Satish Narayanasamy.
\newblock Chimera: hybrid program analysis for determinism.
\newblock In {\em {ACM} {SIGPLAN} Conference on Programming Language Design and
  Implementation (PLDI)}, pages 463--474, 2012.

\bibitem{xiao12:infoflow}
Xusheng Xiao, Nikolai Tillmann, Manuel Fahndrich, Jonathan De~Halleux, and
  Michal Moskal.
\newblock User-aware privacy control via extended static-information-flow
  analysis.
\newblock In {\em International Conference on Automated Software Engineering
  (ASE)}, pages 80--89, 2012.

\bibitem{functionsummary2}
Ben-Chung Cheng and Wen-Mei~W. Hwu.
\newblock Modular interprocedural pointer analysis using access paths: Design,
  implementation, and evaluation.
\newblock In {\em Proceedings of the ACM SIGPLAN 2000 Conference on Programming
  Language Design and Implementation (PLDI)}, page 57–69, 2000.

\bibitem{solidityparsing}
Federico Bond.
\newblock A solidity parser for js built on top of a robust antlr4 grammar,
  2019.
\newblock https://github.com/federicobond/solidity-parser-antlr.

\bibitem{antlr}
Terence Parr.
\newblock {ANTLR}, 2014.
\newblock http://www.antlr.org/.

\bibitem{Bigquery}
Google bigquery, 2021.
\newblock https://cloud.google.com/bigquery/.

\bibitem{tuffle}
Tuffle, 2021.
\newblock https://www.trufflesuite.com/.

\bibitem{multitest1}
Kunal Taneja, Nuo Li, Madhuri~R. Marri, Tao Xie, and Nikolai Tillmann.
\newblock Mitv: multiple-implementation testing of user-input validators for
  web applications.
\newblock In {\em {IEEE/ACM} International Conference on Automated Software
  Engineering (ASE)}, pages 131--134, 2010.

\bibitem{testingbook}
Paul Ammann and Jeff Offutt.
\newblock {\em Introduction to Software Testing}.
\newblock Cambridge University Press, 2008.

\bibitem{gasprice}
Gas and fees, 2021.
\newblock https://ethereum.org/en/developers/docs/gas/.

\bibitem{stackdepth}
Ethereum.
\newblock Solidity security considerations, 2019.
\newblock
  https://solidity.readthedocs.io/en/v0.6.1/security-considerations.html.

\bibitem{rodler2021evmpatch}
Michael Rodler, Wenting Li, Ghassan~O Karame, and Lucas Davi.
\newblock Evmpatch: timely and automated patching of ethereum smart contracts.
\newblock In {\em 30th $\{$USENIX$\}$ Security Symposium ($\{$USENIX$\}$
  Security 21)}, 2021.

\bibitem{move}
Facebook.
\newblock Move: A language with programmable resources, 2019.
\newblock
  https://developers.libra.org/docs/assets/papers/libra-move-a-language-with-programmable-resources.pdf.

\bibitem{karame2016bitcoin}
Ghassan~O Karame and Elli Androulaki.
\newblock {\em Bitcoin and blockchain security}.
\newblock Artech House, 2016.

\bibitem{ekiden}
Raymond Cheng, Fan Zhang, Jernej Kos, Warren He, Nicholas Hynes, Noah Johnson,
  Ari Juels, Andrew Miller, and Dawn Song.
\newblock Ekiden: A platform for confidentiality-preserving, trustworthy, and
  performant smart contracts.
\newblock In {\em 2019 IEEE European Symposium on Security and Privacy
  (EuroS\&P)}, pages 185--200. IEEE, 2019.

\bibitem{solidus}
Ethan Cecchetti, Fan Zhang, Yan Ji, Ahmed Kosba, Ari Juels, and Elaine Shi.
\newblock Solidus: Confidential distributed ledger transactions via pvorm.
\newblock In {\em Proceedings of the 2017 ACM SIGSAC Conference on Computer and
  Communications Security}, pages 701--717, 2017.

\bibitem{kosba2016hawk}
Ahmed Kosba, Andrew Miller, Elaine Shi, Zikai Wen, and Charalampos Papamanthou.
\newblock Hawk: The blockchain model of cryptography and privacy-preserving
  smart contracts.
\newblock In {\em 2016 IEEE symposium on security and privacy (SP)}, pages
  839--858. IEEE, 2016.

\bibitem{callback}
Shelly Grossman, Ittai Abraham, Guy Golan-Gueta, Yan Michalevsky, Noam
  Rinetzky, Mooly Sagiv, and Yoni Zohar.
\newblock Online detection of effectively callback free objects with
  applications to smart contracts.
\newblock {\em Proceedings of the ACM on Programming Languages}, 2(POPL):48,
  2017.

\bibitem{apr}
Claire {Le Goues}, Michael Pradel, and Abhik Roychoudhury.
\newblock Automated program repair.
\newblock {\em Commun. {ACM}}, 62(12):56--65, 2019.

\end{thebibliography}
